\documentclass[12pt,a4paper]{article}
\usepackage[a4paper, total={6in,9in}]{geometry}

\usepackage{amsmath}
\usepackage{amsfonts}
\usepackage{amssymb}
\usepackage{graphicx}
\usepackage{color}
\usepackage{epstopdf}
\usepackage{xcolor,colortbl}
\usepackage{caption}
\usepackage{tikz}
\usepackage{subfigure}

 \usepackage{ulem}
 
\begin{document}

\begin{center}
\begin{large}
\textbf{MicroMegascope based dynamic Surface Force Apparatus}
\end{large}
\end{center}

\begin{center}
Antoine Lain\'e$^1$, Laetitia Jubin$^1$, Luca Canale$^1$, Lyd\'eric Bocquet$^1$, Alessandro Siria$^1$, Stephen H. Donaldson Jr$^2$ and Antoine Nigu\`es$^1$\\
\vspace{0.2cm}
$^1$ \textit{Laboratoire de Physique Statistique de l'\'Ecole Normale Sup\'erieure, ENS, Universit\'e PSL, CNRS, Sorbonne Universit\'e, Universit\'e Paris-Diderot, Sorbonne Paris Cit\'e, UMR CNRS 8550, 24 Rue Lhomond 75005 Paris, France}\\
$^2$ \textit{D\'epartement de Physique, Ecole Normale Sup\'erieure - PSL Research University, CNRS, 24 rue Lhomond 75005 Paris, France}
\end{center}

Keywords : Nanorheology, Nanotribology, Surface Forces, Tuning Fork

\begin{abstract}


Surface Force Apparatus (SFA) allows to accurately resolve the interfacial properties of fluids confined between extended surfaces. The accuracy of the SFA makes it an ubiquitous tool for the nanoscale mechanical characterization of soft matter systems. 
The SFA traditionally measures force-distance profiles through interferometry with subnanometric distance precision. However, these techniques often require a dedicated and technically demanding experimental setup, and there remains a need for versatile and simple force-distance measurement tools.
Here we present a MicroMegascope based dynamic Surface Force Apparatus capable of accurate measurement of the dynamic force profile of a liquid confined between a millimetric sphere and a planar substrate. Normal and shear mechanical impedance is measured within the classical Frequency Modulation framework. We measure rheological and frictional properties from micrometric to molecular confinement. We also highlight the resolution of small interfacial features such as ionic liquid layering. This apparatus shows promise as a versatile force-distance measurement device for exotic surfaces or extreme environments.

\end{abstract}

\clearpage
\section{Introduction}

The Surface force apparatus (SFA) is an invaluable instrument to resolve surface forces in a wide variety of soft matter systems. Indeed, probing the mechanical properties of highly confined soft or fluidic systems is of great interest in the fundamental understanding of many research fields such as nanofluidics \cite{Bocquet2009}, lubrication \cite{Persson} and energy storage \cite{Fedorov2014}. The classical SFA was originally developed for measuring equilibrium surface forces \cite{Israelachvili1978}, and then several design updates included the ability to probe frictional properties \cite{Gee1990, Klein1191, Israelachvili2010}. More recently, dynamic SFA highlights the non-contact rheological responses of confined liquids \cite{Garcia2016}. 
Classically, a SFA measures the surface interaction across a liquid confined between two extended surfaces. The surface forces are measured by mounting the lower surface on a cantilever spring. The upper surface is mounted on a positional actuator and its displacement induces a confinement change. The force is computed from the stiffness of the spring and the difference between the imposed and actual separation distance \cite{Israelachvili2010}. Thus the distance between the two surfaces has to be accurately measured in order to extract the force profile. Several techniques such as interferometric or capacitive measurements were developed in order to reach a subnanometric resolution on the distance measurement \cite{Israelachvili1973, Restagno2001, Restagno2002}. Recently, dynamic SFA enabled hydrodynamic force measurement superimposed to the static response \cite{Garcia2016}. The deflection of mechanical part along with the imperative need of confinement distance measurement brings experimental complexity and imposes constraints on the surface design and environment.\\ 
Here, we present a versatile macroscopic tuning fork based apparatus capable of probing the two-dimensional - normal and tangential - mechanical impedance of confined liquid between extended surfaces by dynamic force measurement. The working principle of the force sensor is similar to the MicroMegascope \cite{Canale2018}, where the probe is now a millimetric sphere. The mechanical response resulting from the interaction between the sphere immersed in liquid and the substrate is obtained from classical Frequency Modulation signal treatment technique \cite{RevModPhys.75.949}. Such sensor allies high force resolution along with high stiffness. As a consequence the entire setup presents a very high stiffness which allows for reduction of effects due to mechanical instabilities (snap-in, adhesion hysteresis) and ensures that no significant mechanical deformations occur. Therefore the surface separation is imposed and not measured. 
To illustrate the capabilities of the apparatus, we present two-dimensional dynamic force measurements, which reveals both rheological and frictional properties of the system as the probe-substrate distance is varied. Additionally, we show the resolution of ionic liquid layering for nanometer-size confinement.

\section{Experimental setup}

\begin{figure}[htb!]
\centering
\includegraphics[width=\columnwidth]{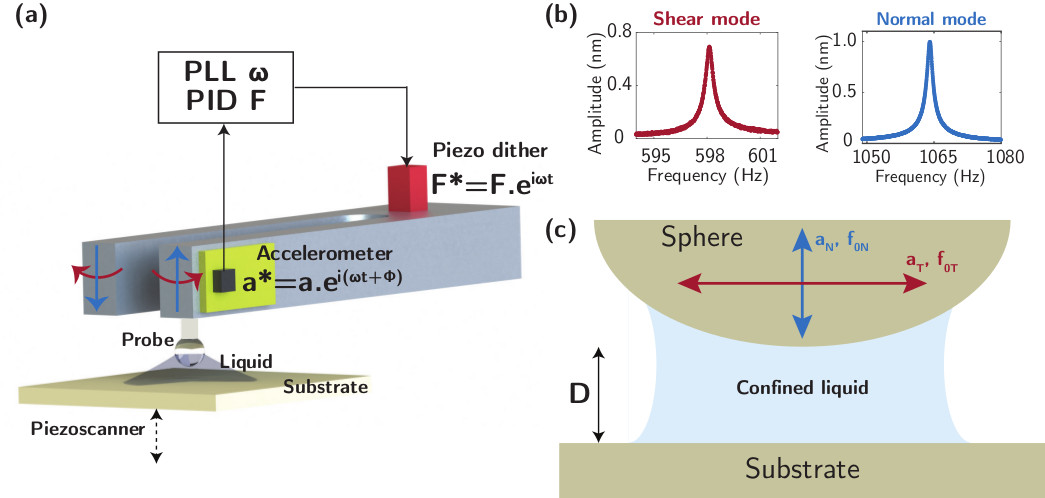}
\caption{\textbf{Experimental setup} \textbf{(a)} Schematical representation of the experimental setup. The liquid is confined between the substrate and the spherical probe which radius of curvature is $R \approx 1.5$ mm. A piezoscanner controls the confinement distance $D$ over tens of micrometers with a subnanometric precision. A piezodither mechanically excites the tuning fork $F^* = F \cdot e^{i \omega t}$ [N]. A two-axis accelerometer records the tangential and normal amplitude $a^* = a \cdot e^{(i \omega t + \phi)}$ [m]. For both oscillation directions, a Phase-Locked Loop (PLL) dynamically tracks the frequency modulation ($\omega$) to remain at the system resonance and a PID retroactively adjusts the intensity ($F$) of the excitation signal in order to keep the oscillation amplitude constant. \textbf{(b)} Unperturbed resonance curves for the two oscillation modes are displayed. The tangential resonance curve is centered on $f_{0T} \approx 598$ Hz (red curve) and the normal mode on $f_{0N} \approx 1060$ Hz (blue curve). \textbf{(c)} Zoom on the sphere-plane confinement geometry. The spherical probe oscillates with a normal amplitude $a_N$ and shear amplitude $a_T$. The substrate displacement imposes the confinement distance D.}
\label{fig:exp-setup}
\end{figure}


We show in figure \ref{fig:exp-setup}\textbf{(a)} a schematic representation of the experimental setup. 
The plane surface is mounted on a fully calibrated piezoscanner (PiezoJena MicroTRITOR) which accurately controls the substrate displacement over tens of micrometers with subnanometric precision. The absolute position of the zero distance is defined from the hard contact observed in the friction response, neglecting the surfaces deformation resulting from Hertzian contact, and then the distance change is only due to the piezoscanner expansion as the entire setup is fully rigid. The approach velocity can be controlled in the range 0.01 up to 100 nm.s$^{-1}$. The piezoscanner stiffness lies above $10^6$ N.m$^{-1}$ and therefore does not experience any deflection. A glass rod with a spherical end of millimetric radius (R $\approx 1.5$ mm) is attached to one prong of a centimetric tuning fork. The liquid is then confined in a sphere-plane geometry as illustrated in figure \ref{fig:exp-setup}\textbf{(c)}.\\
Similarly to the MicroMegascope \cite{Canale2018}, the force sensor is a centimetric tuning fork. Here, the prongs of the tuning fork are 10 cm long, 15 mm high and 7.5 mm thick. The device is especially designed to present orthogonal resonance modes as previously performed with quartz tuning fork \cite{Comtet2017corn}. Figure \ref{fig:exp-setup}\textbf{(b)} shows the two resonance curves for the shear mode ($\approx 600$ Hz) and the normal mode ($\approx 1000$ Hz). Both modes exhibit a very high stiffness above $10^5$ N.m$^{-1}$ and a quality factor of the order of thousands. Several advantages arise from the high normal stiffness ($K_N \approx 9 \cdot 10^5$ N.m$^{-1}$) : it prevents any significant probe deflection (as a 1 uN force would imply a deflection $\delta < 10$ pm) or mechanical instability, and enables one to probe very stiff systems. The minimal force detection for both modes is of the order of 30 and 50 pN/$\sqrt{\text{Hz}}$ for the shear and normal mode respectively \cite{Chaste2012}. Finally, the large range of oscillation amplitude achievable by this macroscopic tuning fork, from hundreds of nanometers down to hundreds of picometers, gives the opportunity to explore a wide range of applied strains. \\ 
In order to dynamically control the normal and shear oscillation, a piezodither (AE050 5D16F from Thorlabs) mechanically excites both oscillation modes of the tuning fork $F^* = F \cdot e^{i \omega t}$ [N]. A two-axis accelerometer (LIS344ALHTR from STMicroelectronics) monitors the resulting normal and tangential oscillation amplitudes $a^* = a \cdot e^{i(\omega t + \phi)}$ where $\phi$ is the phase difference between the excitation and the oscillation signal. For each oscillation mode a PLL tracks the resonance frequency ($\omega$) and a PID actively controls the excitation input ($F$) to keep the oscillation amplitude constant. The electronic control is performed with a Nanonis controller which possesses its own data treatment software and force spectroscopy module. Such a device enables one to access the mechanical impedance of the probed system $Z^* = F^*/a^* = Z' + i Z''$ [N.m$^{-1}$], defined as the ratio between the dynamic force $F^*$ and the dynamic oscillation amplitude $a^*$. The apparatus then probes the response of the system under dynamic imposed strain.

 
The features of the resonance are modified when the probe starts interacting with its environment. The frequency shift ($\delta f$) is directly related to the conservative force response $Z'$ in phase with the oscillation : 
\begin{equation}
Z' = K_0 \cdot \left( \frac{2\delta f}{f_0} + \left( \frac{\delta f}{f_0} \right)^2 \right) 
\end{equation}
where $K_0$ is the mode stiffness and $f_0$ the bare resonance frequency. $Z'$ corresponds to a measure of the elasticity of the probed system. The ratio $\delta f / f_0$ reaches $\approx 10\%$ for strong confinement at which point the conservative impedance computation must include the second order terms that are usually neglected in Frequency Modulation analysis \cite{RevModPhys.75.949}. The broadening of the response is a signature of a dissipative loss in the system and relates to the imaginary part of the mechanical impedance $Z''$ as :
\begin{equation}
Z'' = K_0 \cdot \left( \frac{1}{Q} - \frac{1}{Q_0} \right) 
\end{equation}
where $Q_0$ is the non-interacting quality factor and $Q$ the quality factor while interactions occur. The low intrinsic dissipation, highlighted by quality factor as high as thousands while the probe is immersed in liquid, enables one to explore low-dissipative phenomena.\\
We show in the following that the apparatus performs quantitative impedance measurement of both long-range hydrodynamic and surface forces. \\

\section{Results}
\subsection{Nanorheology and Nanotribology measurements}

\begin{figure}[htb!]
\centering
\includegraphics[width=\columnwidth]{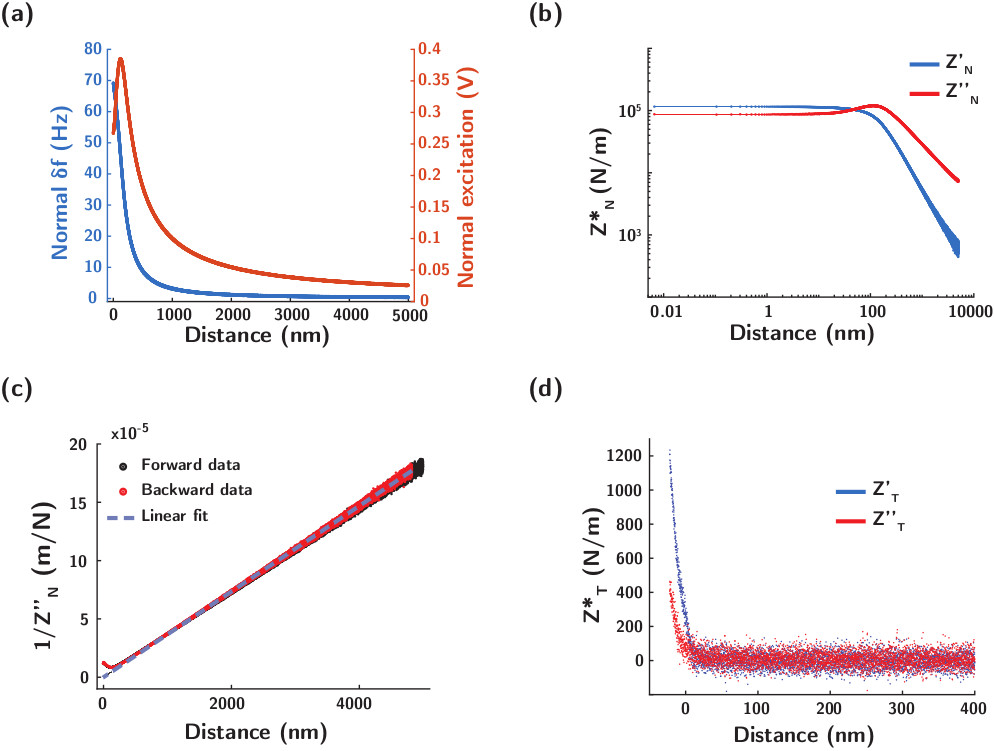}
\caption{\textbf{Nanorheology/Nanotribology measurement} : for Silicone oil confined between a Mica substrate and a glass sphere \textbf{(a)} Raw frequency shift (left axis) and excitation (right axis) obtained by active retroaction on the normal oscillation amplitude as a function of confinement distance. The oscillation amplitude is monitored at $a_N \approx 500$ pm. The approach velocity is $v_Z \approx 0.2$ nm/s. \textbf{(b)} Normal mechanical impedance $Z^*_N$ obtained from the raw data of (a). \textbf{(c)} Inverse of the imaginary normal impedance as a function of distance. Forward(resp. backward) data corresponds to a decreasing(resp increasing) confinement distance. The thin lines represents the linear fit of the drainage viscous force. The correspondance of the forward and backward responses discard any hysteresis and points out the stability of the system. \textbf{(d)} Tangential mechanical impedance $Z^*_T$ as a function of distance. The tangential amplitude is $a_T \approx 500$ pm.}
\label{fig:rheo} 
\end{figure}

In order to illustrate the capabilities of the apparatus we present the rheological response of silicone oil (Sigma Aldrich, Silicone oil AP 100) confined between a glass sphere and a freshly cleaved Mica substrate. 

Figure \ref{fig:rheo}\textbf{(a)} presents the raw data obtained for the normal oscillation mode while varying the confinement distance from several micrometers down to the solid contact. The normal impedance $Z^*_N$ shown in figure \ref{fig:rheo}\textbf{(b)} is computed from the raw data using the framework detailed in the previous section. While decreasing the separation distance we observe a cross-over from a long-range mainly dissipative regime to a stiffness-dominated regime.

A Newtonian fluid confined by infinitely rigid surfaces in a sphere-plane geometry with no-slip boundary condition, presents a purely dissipative mechanical impedance in response to an harmonically driven normal oscillatory strain. Then assuming that the confinement distance follows $D \ll R$ the dissipative impedance reads \cite{Garcia2016} : 
\begin{equation}
Z''_N = \frac{6 \pi \eta R^2 \cdot 2 \pi f_{0N}}{D}
\end{equation}

with $\eta$ the fluid viscosity, $R$ the sphere radius, $f_{0,N}$ the normal oscillation frequency. Figure \ref{fig:rheo}\textbf{(c)} shows the plot of 1/Z'' versus distance. The linear relationship from several micrometers down to hundreds of nanometers shows that the normal damping is well described by the viscous drainage force and the viscosity remains constant at its bulk value over this whole confinement range. We infer from the slope a viscosity $\eta \approx 96$ mPa.s in excellent agreement with the tabulated value of $100$ mPa.s.\\ 
For confinement below hundreds of nanometers, we observe a systematic deviation from the drainage dissipative response along with a stiffness increase that we attribute to elastohydrodynamic effects as previously observed in dynamic SFA \cite{Garcia2016} with smaller frequencies. In the elastohydrodynamic framework, under a critical confinement distance, part of the liquid is clamped by its viscosity and the mechanical response becomes dominated by the surface deflection \cite{Leroy2011, Villey2013} then giving access to the surface mechanical properties \cite{Leroy2012} without contact. Our results present a good qualitative agreement with the elastohydrodynamic model.\\
This device enables simultaneous measurement of the bulk rheological properties and the tribological response of the system. Along with the normal drainage response, figure \ref{fig:rheo}\textbf{(d)} presents the tangential impedance $Z^*_T$ of the system as a function of separation distance. Far from the substrate, we observe no significant tangential response. For small oscillation amplitudes, the associated shear rate $\dot{\gamma} = \frac{a_T 2 \pi f_{0T}}{D}$ remains small and the associated shear viscous force is negligible. Then a sharp increase of the tangential stiffness and dissipation is observed when the confinement surfaces come into direct contact. Therefore we use this hard contact frictional response in order to define the absolute position of the probe-substrate contact. In practice, during a spectroscopy experiment, the displacement of the substrate relatively to the probe is varied by applying a voltage drop to the piezoscanner. The high stiffness of the whole setup ensures that the piezoscanner expansion corresponds directly to the separation distance decrease. Then, after the approach-retract experiment is performed, the position of the zero separation distance is defined by translation in order to match the one inferred from the frictional response.\\
We observe that the tangential stiffness dominates the mechanical response for such small shear amplitude $a_T = 500$ pm which is reminiscent of hard solid contact. For higher shear rate, one can expect lubricated contact to drive the friction response of the system.

%

%




\subsection{Surface forces and interfacial layering}

\begin{figure}[htb!]
\centering
\includegraphics[width=0.9\columnwidth]{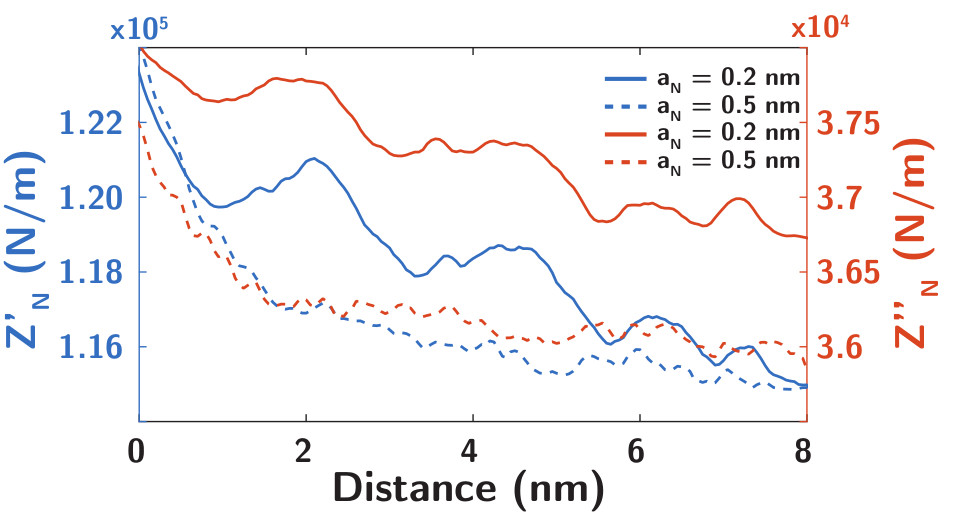}
\caption{\textbf{Molecular forces} : Molecular layering of ionic liquid between a mica and glass surface. The plot displays the normal mechanical impedance $Z^*_N$ as a function of confinement distance. The stiffness $Z'_N$ (left axis, blue curves) and dissipative impedance $Z''_N$ (right axis, orange curves) are measured for amplitudes of 0.2 nm (solid curves) and 0.5 nm (dashed curves). The experiments are performed in ambient conditions.}
\label{fig:layer}
\end{figure}


To emphasize the various capabilites of the apparatus, we present measurement of near-contact surface forces in addition to the long-range nanorheological measurements presented above. Figure \ref{fig:layer} shows the conservative and dissipative mechanical impedance measured over a $D \approx 8$ nm gap across an ionic liquid (Bmim PF6, Sigma Aldrich) confined between the glass bead and a flat mica surface. Progressive peaks in both the conservative and dissipative response with a periodicity between $0.5-1$ nm are measured as the distance decreases, indicative of a structural force in which layers are progressively squeezed out from the contact zone. We note that this periodicity is in good agreement with the ionic liquid typical molecular size of $l \approx 0.7$ nm \cite{Li2013}. Over a representative sample of 10 separate force measurements, we observe an average of 7 peaks. \\

Clear correlations between the conservative and dissipative forces are observed. The peaks in the dissipative force occur about $2-3$ \AA$~$after (i.e., at a smaller distance than) the correlated peak in the conservative force. This corresponds to the stiffness (conservative force) increasing during the compression of a layer, followed by a sharp decrease in the stiffness when the layer is finally squeezed out. The dissipative force, on the other hand, reaches its peak just after the layer begins to be squeezed out, corresponding to a maximum in the layer viscosity, and then drops during the squeeze out, as the layer viscosity decreases. The choice for the oscillation amplitude $a = 0.2$ nm results from an interplay between reducing the noise on the force signal by increasing the amplitude and keeping it smaller than the physical features of the system at stake, here $l \approx 0.7$ nm. The oscillatory force profile is suppressed when the oscillation amplitude of the tuning fork is increased to 0.5 nm, indicating that the layering is disrupted when the amplitude of the tuning fork oscillation approaches the size of the molecular layers. \\

These measurements are reminiscent of structural forces measured by SFA and AFM, which have been shown to exhibit similar oscillatory behaviors in ionic liquid systems \cite{Smith2013, Garcia2017, Perkin2012, Bou-Malham2010, Ueno2010}. Both the onset distance ($D \approx 7$ nm) and number of observed layers ($\approx$ 7) correspond closely with previous SFA measurements on the same ionic liquid \cite{Bou-Malham2010}. Most previous measurements were done in a symmetric mica-mica surface configuration. The consistency between the present measurements in the asymmetric glass-mica configuration and the previous measurements indicate that the liquid structure adjacent to the glass surface is not significantly different compared to the mica surface, likely since both glass and mica are anionic surfaces \cite{Bou-Malham2010}. Therefore, this macro tuning-fork based system allows measurement of structural layering forces while providing simultaneous insight into the dissipative viscous forces operative during layering transitions.


\section{Conclusion}

In summary, we observe that our MicroMegascope based dynamic Surface Force Apparatus is capable of measuring the micrometer-range rheological response as well as the surface forces down to molecular confinement. Probing both the normal and tangential mechanical impedance gives access to the rheological and tribological properties of the confined liquid. 
The apparatus is able to reveal with unprecedented ease the effect of imposed shear on the rheology of nanoconfined complex fluids in a similar spirit to previous AFM experiments \cite{li2014} with small nanometric probes. Furthermore, it is a system of choice to study the elastohydrodynamic lift force resulting from a shear velocity and mediated by the surface deformation close to contact \cite{Skotheim2005, Urzay2007}.\\
Our apparatus is well suited for the study of tribological phenomena. In fact, nowadays there is an increasing appeal to bridge the AFM nanoscale friction experiments to the macroscopic multi-asperities friction measurements. Recently, Garabedian et al. \cite{Garabedian2019} developped a new method to use conventional AFM cantilever to perform high force measurement with colloidal probe. The high sensitivity along with the high stiffness of our apparatus enable to measure forces from the nanoscale up to the macroscale with relative easiness.

The confinement distance is directly imposed, thus removing the need of a technically demanding distance measurement setup. Such simplification enables versatility, for example, facile use of the apparatus under vacuum, and allows a wide variety of surfaces and liquids to be examined.
The use of metallic surfaces would enable studying the effects of electrostatic interactions on the rheological and tribological properties of confined liquids \cite{Fedorov2014, Fajardo2015}.

\section*{Aknowledgements}

A.S. acknowledges funding from the European Union’s H2020 Framework Programme/ ERC Starting Grant agreement number 637748 - NanoSOFT. L.B. acknowledges funding from the European Union’s H2020 Framework Programme/ERC Advanced Grant Shadoks. SHD was supported by LabEX ENS-ICFP: ANR-10-LABX-0010/ANR-10-IDEX-0001-02 PSL*.



\bibliographystyle{ieeetr}
\bibliography{TFdSFA}

\begin{thebibliography}{10}

\bibitem{Bocquet2009}
L.~Bocquet and E.~Charlaix, ``{Nanofluidics, from bulk to interfaces},'' {\em
  Chemical Society Reviews}, vol.~39, pp.~1073--1095, 2009.

\bibitem{Persson}
B.~N.~J. Persson, {\em Sliding Friction: Physical Principles and Applications}.
\newblock 2000.

\bibitem{Fedorov2014}
M.~V. Fedorov and A.~A. Kornyshev, ``{Ionic Liquids at Electrified
  Interfaces},'' {\em Chemical Reviews}, vol.~114, pp.~2978--3036, 2014.

\bibitem{Israelachvili1978}
J.~Israelachvili and G.~E. Adams, ``{Measurement of Forces between Two Mica
  Surfaces in Aqueous Electrolyte Soutions in the Range 0-100 nm.},'' {\em
  J.Chem.Soc., Faraday Trans.I}, vol.~74, p.~975, 1978.

\bibitem{Gee1990}
M.~L. Gee, P.~M. McGuiggan, J.~N. Israelachvili, and A.~M. Homola, ``{Liquid to
  solidlike transitions of molecularly thin films under shear},'' {\em The
  Journal of Chemical Physics}, vol.~93, no.~3, pp.~1895--1906, 1990.

\bibitem{Klein1191}
J.~Klein, D.~Perahia, and S.~Warburg, ``{Forces between polymer-bearing
  surfaces undergoing shear},'' {\em Nature}, vol.~352, pp.~143--145, 1191.

\bibitem{Israelachvili2010}
J.~Israelachvili, Y.~Min, M.~Akbulut, A.~Alig, G.~Carver, W.~Greene,
  K.~Kristiansen, E.~Meyer, N.~Pesika, K.~Rosenberg, and H.~Zeng, ``{Recent
  advances in the surface forces apparatus (SFA) technique},'' {\em Reports on
  Progress in Physics}, vol.~73, no.~3, 2010.

\bibitem{Garcia2016}
L.~Garcia, C.~Barraud, C.~Picard, J.~Giraud, E.~Charlaix, and B.~Cross, ``{A
  micro-nano-rheometer for the mechanics of soft matter at interfaces},'' {\em
  Review of Scientific Instruments}, vol.~87, no.~11, 2016.

\bibitem{Israelachvili1973}
J.~Israelachvili, ``Thin film studies using multiple-beam interferometry,''
  {\em Journal of Colloid and Interface Science}, vol.~44, no.~2, pp.~259 --
  272, 1973.

\bibitem{Restagno2001}
F.~Restagno, J.~Crassous, E.~Charlaix, and M.~Monchanin, ``A new capacitive
  sensor for displacement measurement in a surface-force apparatus,'' {\em
  Measurement Science and Technology}, vol.~12, no.~1, p.~16, 2001.

\bibitem{Restagno2002}
F.~Restagno, J.~Crassous, Ã.~Charlaix, C.~Cottin-Bizonne, and M.~Monchanin, ``A
  new surface forces apparatus for nanorheology,'' {\em Review of Scientific
  Instruments}, vol.~73, no.~6, pp.~2292--2297, 2002.

\bibitem{Canale2018}
L.~Canale, A.~Laborieux, A.~A. Mogane, L.~Jubin, J.~Comtet, A.~Lain{\'{e}},
  L.~Bocquet, A.~Siria, and A.~Nigu{\`{e}}s, ``Micromegascope,'' {\em
  Nanotechnology}, vol.~29, no.~35, p.~355501, 2018.

\bibitem{RevModPhys.75.949}
F.~J. Giessibl, ``Advances in atomic force microscopy,'' {\em Rev. Mod. Phys.},
  vol.~75, pp.~949--983, Jul 2003.

\bibitem{Comtet2017corn}
J.~Comtet, G.~Chatt{\'{e}}, A.~Nigu{\`{e}}s, L.~Bocquet, A.~Siria, and
  A.~Colin, ``{Pairwise frictional profile between particles determines
  discontinuous shear thickening transition in non-colloidal suspensions},''
  {\em Nature Communications}, vol.~8, pp.~1--17, 2017.

\bibitem{Chaste2012}
J.~Chaste, A.~Eicher, J.~Moser, R.~Rurali, and A.~Bachtold, ``A nanomechanical
  mass sensor with yoctogram resolution,'' {\em Nature Nanotechnology}, vol.~7,
  pp.~301 -- 304, 2012.

\bibitem{Leroy2011}
S.~Leroy and E.~Charlaix, ``{Hydrodynamic interactions for the measurement of
  thin film elastic properties},'' {\em Journal of Fluid Mechanics}, vol.~674,
  pp.~389--407, 2011.

\bibitem{Villey2013}
R.~Villey, E.~Martinot, C.~Cottin-Bizonne, M.~Phaner-Goutorbe, L.~L\'eger,
  F.~Restagno, and E.~Charlaix, ``Effect of surface elasticity on the rheology
  of nanometric liquids,'' {\em Phys. Rev. Lett.}, vol.~111, p.~215701, Nov
  2013.

\bibitem{Leroy2012}
S.~Leroy, A.~Steinberger, C.~Cottin-Bizonne, F.~Restagno, L.~L\'eger, and
  E.~Charlaix, ``Hydrodynamic interaction between a spherical particle and an
  elastic surface: A gentle probe for soft thin films,'' {\em Phys. Rev.
  Lett.}, vol.~108, p.~264501, Jun 2012.

\bibitem{Li2013}
H.~Li, F.~Endres, and R.~Atkin, ``{Effect of alkyl chain length and anion
  species on the interfacial nanostructure of ionic liquids at the
  Au(111)–ionic liquid interface as a function of potential},'' {\em Physical
  Chemistry Chemical Physics}, vol.~15, no.~35, p.~14624, 2013.

\bibitem{Smith2013}
A.~M. Smith, K.~R.~J. Lovelock, N.~N. Gosvami, T.~Welton, and S.~Perkin,
  ``{Quantized friction across ionic liquid thin films},'' {\em Physical
  Chemistry Chemical Physics}, vol.~15, no.~37, p.~15317, 2013.

\bibitem{Garcia2017}
L.~Garcia, L.~Jacquot, E.~Charlaix, and B.~Cross, ``{Nano-mechanics of ionic
  liquids at dielectric and metallic interfaces},'' {\em Faraday Discuss.},
  2017.

\bibitem{Perkin2012}
S.~Perkin, ``Ionic liquids in confined geometries,'' {\em Phys. Chem. Chem.
  Phys.}, vol.~14, pp.~5052--5062, 2012.

\bibitem{Bou-Malham2010}
I.~Bou-Malham and L.~Bureau, ``Nanoconfined ionic liquids: effect of surface
  charges on flow and molecular layering,'' {\em Soft Matter}, vol.~6,
  pp.~4062--4065, 2010.

\bibitem{Ueno2010}
K.~Ueno, M.~Kasuya, M.~Watanabe, M.~Mizukami, and K.~Kurihara, ``Resonance
  shear measurement of nanoconfined ionic liquids,'' {\em Phys. Chem. Chem.
  Phys.}, vol.~12, pp.~4066--4071, 2010.

\bibitem{li2014}
T.-d. Li, H.-c. Chiu, D.~Ortiz-young, and E.~Riedo, ``{Nanorheology by atomic
  force microscopy},'' {\em Review of Scientific Instrument}, vol.~123707,
  no.~December, pp.~1--7, 2014.

\bibitem{Skotheim2005}
J.~M. Skotheim and L.~Mahadevan, ``{Soft lubrication : The elastohydrodynamics
  of nonconforming and conforming contacts},'' {\em Physics of fluids},
  vol.~17, pp.~1--23, 2005.

\bibitem{Urzay2007}
J.~Urzay, S.~G. {Llewellyn Smith}, and B.~J. Glover, ``{The elastohydrodynamic
  force on a sphere near a soft wall},'' {\em Physics of Fluids}, vol.~19,
  no.~10, pp.~1--7, 2007.

\bibitem{Garabedian2019}
N.~T. Garabedian, H.~S. Khare, R.~W. Carpick, and D.~L. Burris, ``Afm at the
  macroscale: Methods to fabricate and calibrate probes for millinewton force
  measurements,'' {\em Tribology Letters}, vol.~67, p.~21, Jan 2019.

\bibitem{Fajardo2015}
O.~Y. Fajardo, F.~Bresme, A.~A. Kornyshev, and M.~Urbakh, ``{Electrotunable
  Friction with Ionic Liquid Lubricants: How Important Is the Molecular
  Structure of the Ions?},'' {\em Journal of Physical Chemistry Letters},
  vol.~6, no.~20, pp.~3998--4004, 2015.

\end{thebibliography}

\end{document}